\documentclass[sigconf,natbib=false]{acmart}
\usepackage{enumitem}
\usepackage{tcolorbox}
\usepackage{tabularx}
\usepackage{graphicx}
\usepackage{color}
\usepackage{multirow}

\AtBeginDocument{%
  }

\setcopyright{acmlicensed}
\copyrightyear{2018}
\acmYear{2018}
\acmDOI{XXXXXXX.XXXXXXX}
\acmConference[Conference acronym 'XX]{Make sure to enter the correct
  conference title from your rights confirmation email}{June 03--05,
  2018}{Woodstock, NY}
\acmISBN{978-1-4503-XXXX-X/2018/06}

\RequirePackage[
  datamodel=acmdatamodel,
  style=acmnumeric,
  ]{biblatex}

\begin{document}

\title{The 3rd Place Solution of CCIR CUP 2025: A Framework for Retrieval-Augmented Generation in Multi-Turn Legal Conversation}

\author{Da Li}
\email{lida.ucas@gmail.com}
\affiliation{
	\institution{CAS Key Lab of Network Data}
        \institution{Science and Technology, ICT, CAS}
        \institution{State Key Laboratory of AI Safety}
	\institution{University of Chinese Academy of Sciences}
	\city{Beijing}
	\country{China}
}

\author{Zecheng Fang}
\email{fangzecheng20z@ict.ac.cn}
\affiliation{
    \institution{Institute of Computing Technology, Chinese Academy of Sciences}
	\institution{University of Chinese Academy of Sciences}
	\city{Beijing}
	\country{China}
}

\author{Qiang Yan}
\email{yanqiang@ict.ac.cn}
\affiliation{
	\institution{CAS Key Lab of Network Data}
        \institution{Science and Technology, ICT, CAS}
        \institution{State Key Laboratory of AI Safety}
	\institution{University of Chinese Academy of Sciences}
	\city{Beijing}
	\country{China}
}

\author{Wei Huang}
\email{huangwei21b@ict.ac.cn}
\affiliation{
	\institution{CAS Key Lab of Network Data}
        \institution{Science and Technology, ICT, CAS}
        \institution{State Key Laboratory of AI Safety}
	\institution{University of Chinese Academy of Sciences}
	\city{Beijing}
	\country{China}
}

\author{Xuanpu Luo}
\email{seuviplxp@163.com}
\affiliation{
    \institution{CAS Key Lab of Network Data}
        \institution{Science and Technology, ICT, CAS}
        \institution{State Key Laboratory of AI Safety}
	\institution{University of Chinese Academy of Sciences}
	\city{Beijing}
	\country{China}
}

\renewcommand{\shortauthors}{Da Li et al.}

\begin{abstract}
  Retrieval-Augmented Generation has made significant progress in the field of natural language processing. By combining the advantages of information retrieval and large language models, RAG can generate relevant and contextually appropriate responses based on items retrieved from reliable sources. This technology has demonstrated outstanding performance across multiple domains, but its application in the legal field remains in its exploratory phase. In this paper, we introduce our approach for "Legal Knowledge Retrieval and Generation" in CCIR CUP 2025, which leverages large language models and information retrieval systems to provide responses based on laws in response to user questions.
\end{abstract}



\received{20 February 2007}
\received[revised]{12 March 2009}
\received[accepted]{5 June 2009}

\maketitle

\section{Introduction}
Retrieval-Augmented Generation (RAG) has emerged as a powerful paradigm in natural language processing, combining the strengths of large language models (LLMs) with dynamic knowledge retrieval to enhance the relevance and factual grounding of generated responses. Unlike traditional generative models that rely solely on parametric knowledge, RAG systems integrate real-time information retrieval from external corpora, enabling them to adapt to evolving contexts and domain-specific queries. This approach is particularly valuable in the fields of law and journalism, where users often demand up-to-date information, nuanced reasoning, or specialized knowledge beyond a model's pretrained scope. 

In this paper, we explore the application of RAG to conversation about legal knowledge, demonstrating how retrieval mechanisms—such as dense vector search or keyword-based indexing—can synergize with LLMs like LlaMA~\cite{touvron2023llamaopenefficientfoundation} or Qwen~\cite{yang2025qwen3technicalreport} to improve performance metrics like answer precision, coherence, and diversity. We placed 3rd in the CCIR CUP 2025 competition. We further analyze challenges such as noise filtering and optimal integration strategies, proposing innovations tailored to the competition's unique requirements. Our work underscores RAG's potential to bridge the gap between static pretraining and dynamic real-world demands, offering a scalable solution for knowledge-intensive NLP tasks.

Our work makes several important contributions to the field of legal information retrieval and generation:
\begin{enumerate}[leftmargin=*]
\item \textbf{Conversational Context Handling:} We developed an effective query rewriting mechanism that addresses coreference resolution and topic drift in multi-turn legal conversations.
\item \textbf{Multi-Route Retrieval Strategy:} Our multi-route retrieval approach combines the semantic understanding capabilities of dense retrieval with the precision of keyword-based sparse retrieval, achieving better coverage of relevant legal provisions.
\item \textbf{Efficient Filtering Pipeline:} The article filtering mechanism reduces noise during the retrieval process and provides accurate legal references for generation.
\item \textbf{Legal-Specific Generation:} Our response generation module produces legally accurate and appropriately formatted responses that follow the "syllogistic" reasoning style typical of legal documents.
\end{enumerate}

\section{Related Work}

Large language models (LLMs) based on the Transformer architecture, since the advent of seminal works like BERT~\cite{devlin2019bert} and GPT~\cite{radford2019language}, have become the core technology for addressing knowledge-intensive and multi-hop reasoning tasks. However, the inherent issue of ``hallucination'' in these models limits their applicability in high-stakes domains. To mitigate this, Retrieval-Augmented Generation (RAG) has emerged as a dominant paradigm~\cite{han2024rag, hei2024dr}. By retrieving relevant information from external knowledge sources to ground the generation process, RAG significantly enhances the fidelity of the output. The work presented in this paper is situated within this context, specifically exploring the application of RAG in the legal domain.

In recent years, interdisciplinary research combining RAG with legal applications has burgeoned, leading to a wealth of systematic studies. On one hand, the academic community has begun to establish evaluation standards for Legal-RAG, marked by the introduction of benchmarking suites like LegalBench-RAG~\cite{pipitone2024legalbench} and comprehensive reviews of the state-of-the-art by works such as~\cite{hindi2025enhancing}.
On the other hand, researchers are actively exploring how to integrate the unique structures of legal knowledge into the RAG framework. For instance, studies like Graph-RAG~\cite{de2025graph} and the work by~\cite{barron2025bridging} have investigated leveraging the hierarchical nature of statutes to optimize retrieval. Similarly, CBR-RAG~\cite{wiratunga2024cbr} draws on case-based reasoning to improve question-answering performance by referencing historical judgments. Furthermore, other research focuses on the dynamic optimization of retrieval strategies, such as the adaptive retrieval method proposed in HyPA-RAG~\cite{kalra2024hypa}. These system-level explorations have catalyzed a diverse range of applications, including general-purpose legal assistants~\cite{wahidur2025legal}, dispute resolution solutions for specific legal areas~\cite{rafat2024ai}, and federated RAG architectures that account for data privacy~\cite{amato2024optimizing}, collectively advancing the maturity of Legal-RAG technology.

\section{Preliminary}
\subsection{Pipeline of RAG}
The Retrieval-Augmented Generation (RAG) pipeline consists of three fundamental stages: Retrieval, Augmentation, and Generation. First, the system retrieves relevant documents or data chunks from a knowledge base (e.g., a vector database) based on the user’s query using embedding models and similarity search. Next, the retrieved context is combined with the original query to form an augmented prompt, providing the LLM with grounded information. Finally, the language model generates a coherent and context-aware response using the enriched input. Optionally, pre-processing (e.g., data chunking) and post-processing (e.g., output refinement) can enhance the workflow, but the core RAG mechanism relies on these three key steps to bridge knowledge retrieval with generative AI.

\subsection{Dataset}
We have statistics on the dataset, which are shown in the table below.
\begin{table}[h]
\centering
\caption{Basic statistics of benchmark}
\label{tab:statistics}
\begin{tabular}{cc}
\hline
\textbf{Statistic} & \textbf{\#Number} \\ \hline
Total Conversations & 1,013 \\ 
Total Queries & 5,065 \\ 
Total Legal Articles & 17,229 \\ 
Total Legal Literature & 212 \\
Avg. Query Length & 19.43 \\ 
Avg. Response Length & 165.92 \\ 
Avg. Relevant Articles per Query & 1.09 \\ 
Avg. Keywords per Query & 3.57 \\ \hline
\end{tabular}
\end{table}

\subsection{Evaluation}
To comprehensively assess the performance across the entire task pipeline. Three metrics are involved to evaluate the retrieval result and the response.

\textbf{Retrieval.}
This competition uses NDCG@5 (Normalized Discounted Cumulative Gain) to measure retrieval performance. Normalized Discounted Cumulative Gain is a widely used metric in information retrieval systems to evaluate the quality of search results. It measures how well a ranking algorithm orders items by their relevance, taking into account both the position and the graded relevance of each item in the ranked list. 
\begin{equation}
\centering
\begin{aligned}
Score_{retrieval} = 100 *NDCG@5.
\end{aligned} 
\end{equation}

\textbf{Generation.} 
The generation stage evaluates the quality of the response, focusing on two key aspects: semantic consistency with the reference answer and completeness of critical information. The metrics consist of the following two parts.
\begin{itemize}[leftmargin=*]
    \item BERT-F1: This metric quantifies semantic similarity between the generated response and the reference answer. It leverages a pre-trained BERT model to compute word-level similarity scores, then aggregates these into an F1 score (harmonic mean of precision and recall), reflecting fine-grained semantic alignment.
    \item Keyword\_Accuracy: This metric assesses the coverage of critical information. Given a set of pre-annotated key keywords (essential to the reference answer), it calculates the proportion of these keywords that appear in the generated response, ensuring the retention of core factual content.
\end{itemize}
The generation stage score is computed by equally weighting the two metric:
\begin{equation}
\centering
\begin{aligned}
Score_{generation} = 50*\textit{BERT-F1}+ 50* \textit{Keyword\_Accuracy}.
\end{aligned} 
\end{equation}
To balance the importance of retrieval and generation, the final score is obtained by equally weighting the two stage scores:
\begin{equation}
\centering
\begin{aligned}
Score = 0.5* Score_{retrieval}+ 0.5*Score_{generation}.
\end{aligned} 
\end{equation}
This evaluation framework ensures a holistic assessment, capturing both the retrieval system's ranking capability and the generation model's semantic and factual accuracy.

\begin{figure}[htbp!]
	\centerline{\includegraphics[scale=0.5]{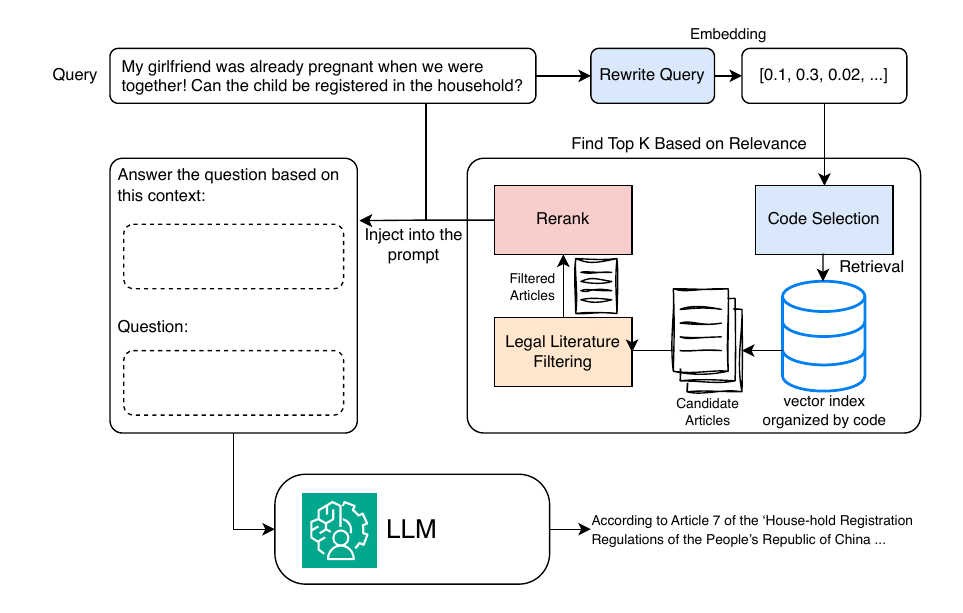}}
	\caption{The pipeline of our method.}
	\label{fig1}
\end{figure}

\section{Methodology}
Based on the vanilla RAG process, we added the following modules to improve the quality of responses. For the competition, participants are required to find legal provisions that can be used to answer user questions. Secondly, they are required to generate responses that are satisfactory to users. We optimized multiple processing steps and generated end-to-end optimized responses.

\subsection{Query Rewriting}
User questions are stored in dialogue format, and historical user questions cannot be ignored when answering each round of user questions. We first use Qwen3 to rewrite user questions and generate queries for the retrieval system.

\begin{figure}[h]
\begin{tcolorbox}[colback=white, colframe=black, sharp corners=southwest, boxrule=0.5pt]
\textbf{Rewriting Prompt:} \\

\textbf{[System]}: You are an expert proficient in legal knowledge, capable of accurately understanding users' questions and identifying users' intentions. Given the following questions (including historical questions and the current question), please rewrite the user's current question into a query for a legal search system to support you in accurately searching legal provisions to answer the user's question. When outputting the corresponding query of the search system, pay special attention to the following requirements:\\
1. Focus on the legal field, and avoid unnecessary wording modifications or the introduction of new terms or concepts not mentioned in the question.\\
2. Summarize or extract the core search intention of the user's current question from the historical questions.\\
...\\
9. Analyze the rewritten result into 3 to 5 key legal terms, and attach the legal terms after the rewritten result. The legal terms must be wrapped in <keyword></keyword> tags, and commas should separate the legal terms. The legal terms need to be directly related to the core legal facts in the rewritten result, and avoid irrelevant terms.\\
\textbf{[User]}: Historical Question: User: Hello, my health certificate has expired, and I was told that I will be fined 10,000 yuan. How should I deal with it? \\
Current Question: I have re-applied for a health certificate and submitted a medical examination report. Can the penalty be reduced or waived?\\
\textbf{[Assistant]}: \textit{Whether re-applying for a health certificate and submitting a medical examination report belong to the circumstances where an administrative penalty can be reduced. <keyword>Re-applying for health certificate, Submitting medical examination report, Reduction of administrative penalty, Timely correction of illegal act, Mitigation of harmful consequences</keyword>.} \\
\end{tcolorbox}
\caption{The prompt of query rewriting. The \textit{italicized text} is the output of the LLM. }
\end{figure}

\subsection{Multi-Route Retrieval}
In legal information retrieval systems, semantic matching techniques demonstrate constrained effectiveness in addressing user queries, which frequently necessitate keyword-dependent matching mechanisms. We adopt a multi-route retrieval strategy. In addition to dense retrieval, we utilize BM25 to expand the candidate set by retrieving relevant legal provisions based on keywords derived from query rewriting.

\begin{figure}[htbp!]
\begin{tcolorbox}[colback=white, colframe=black, sharp corners=southwest, boxrule=0.5pt]
\textbf{Literature Selection:} \\
\textbf{[System]}: You are a helpful assistant.\\
\textbf{[User]}: Given the following user-question and the name of the article, determine whether the user-question may involve the area described by the article. If it is possible, output "Yes", and if not, output "No".\\
Output Format Requirements:\\
1. Any content other than "Yes" or "No" is not allowed in the output.\\
2. Please directly output "Yes" or "No" at the beginning, and do not add any analysis process.\\

User Question:\\
Whether re-applying for a health certificate and submitting a medical examination report belong to the circumstances where an administrative penalty can be reduced?\\

Code Name:\\
Law of the People's Republic of China on Administrative Penalties\\
\textbf{[Assistant]}: \textit{Yes}\\
\end{tcolorbox}
\caption{The prompt of literature selection. The \textit{italicized text} is the output of the LLM. }
\end{figure}

\subsection{Legal Literature Filtering}
Given the applicability of legal articles, there is no need to search the entire corpus for relevant candidates when answering user questions. User questions are only applicable to a small portion of the corpus. We first use LLMs to filter out legal provisions that are irrelevant to user questions. After obtaining a candidate set through multi-path retrieval, we use LLM to filter each legal article in the candidate set secondarily.

\begin{figure}[htbp!]
\begin{tcolorbox}[colback=white, colframe=black, sharp corners=southwest, boxrule=0.5pt]
\textbf{Legal Article Filtering:} \\
\textbf{[System]}: You are a helpful assistant.\\
\textbf{[User]}: Given the following user's question and legal provision, comprehensively judge whether the legal provision is impossible to be used as the key basis for answering the user's question based on the code to which the legal provision belongs (such as the Civil Code of the People's Republic of China) and the content of the legal provision. If yes, output "Yes"; if no, output "No".\\
Output Format Requirements:\\
1. Any content other than "Yes" or "No" is not allowed in the output.\\
2. Please directly output "Yes" or "No" at the beginning, and do not add any analysis process.\\

User Question:\\
Whether re-applying for a health certificate and submitting a medical examination report belong to the circumstances where administrative penalty can be reduced?\\

Literature Name:\\
Article 32 of the Law of the People's Republic of China on Administrative Penalties: A party who is in any of the following circumstances shall be given a lighter or mitigated administrative penalty: ...\\
\textbf{[Assistant]}: \textit{Yes}\\
\end{tcolorbox}
\caption{The prompt of filtering. The \textit{italicized text} is the output of the LLM. }
\end{figure}

\subsection{Reranking}
In retrieval-augmented generation (RAG), reranking serves as a critical intermediate step to address limitations of retrievers and enhance the quality of context fed into LLMs. Because of efficiency, the retrieval phase usually uses a dual-encoder as a retriever to find a candidate set from a huge corpus.
By leveraging sophisticated models (e.g., cross-encoders) that capture bidirectional query-document interactions, reranking refines the initial candidate set through deeper contextual understanding, enabling fine-grained relevance assessment for complex or ambiguous queries. This process reduces noise by filtering low-quality candidates, resolving ranking conflicts to prioritize high-informative documents. Ultimately, reranking directly improves RAG performance by aligning retrieved context with query intent, minimizing hallucinations, and enhancing the factual consistency and coherence of generated outputs. 
As the bottleneck of the RAG system, reranking is closely related to the quality of the response. Based on the training dataset, we retrained a reranking model to learn the relevant matching in the legal field effectively. 

\subsection{Response Generation}
The top 5 legal provisions obtained by reranking are then used as input for the LLM, along with the user's dialogue, to generate the corresponding response. The responses in the dataset exhibit similar styles. To ensure that the LLMs' output maintains the same style, we use in-context learning to provide two response examples in the prompt, thereby ensuring consistency in response style.

\begin{figure}[h]
\begin{tcolorbox}[colback=white, colframe=black, sharp corners=southwest, boxrule=0.5pt]
\textbf{Response Generation} \\
\textbf{[System]}: You are an expert proficient in legal knowledge, dedicated to providing users with accurate and professional legal consultation. Your responses should be rigorous and efficient, and maintain consistency in style with previous rounds of answers (if any). If the user's question involves specific legal provisions, relevant legal clauses should be cited as much as possible to enhance the authority of the answer. When outputting, special attention should be paid to the following requirements:\\
1. Priority should be given to selecting legal provisions from the given reference legal clauses that can be used to answer the user's question as references. Select and cite them in order from the beginning to the end of the reference legal clauses.\\
...\\
5. Please refer to and generate replies similar to the following style.\\
Examples:\\
"According to the provisions of Article 428 of the Civil Code of the People's Republic of China, the vehicle installment payment contract and the vehicle insurance contract should be handled separately. If you repay the monthly payment on time and do not violate the repayment obligations in the installment agreement, the installment company has no right to seize the vehicle. As for insurance purchase, unless the contract clearly stipulates that insurance must be purchased through the installment company, the installment company has no right to seize the vehicle on this ground."\\
\textbf{[User]}:I have re-applied for a health certificate and submitted a medical examination report. Can the penalty be reduced or waived?\\
The following are the legal clauses you can refer to:\\
1. Article 32 of the Law of the People's Republic of China on Administrative Penalties: A party who is in any of the following circumstances shall be given a lighter or mitigated administrative penalty:\\
...\\
\textbf{[Assistant]}: \textit{According to the provisions of Item (1) of Article 32 and Article 33 of the Law of the People's Republic of China on Administrative Penalties, if you have actively re-applied for a health certificate and submitted a medical examination report, you may apply for a reduction or waiver of the penalty in accordance with the law under the following circumstances:...}\\
\end{tcolorbox}
\caption{The prompt of response generation. The \textit{italicized text} is the output of the LLM. }
\end{figure}

\section{Experiments}

\subsection{Experimental Setup}
We conducted comprehensive experiments to evaluate our proposed approach for legal knowledge retrieval and generation. The experimental setup includes the selection of appropriate models for different modules and careful hyperparameter tuning.

\subsubsection{Model Selection}
We employed different specialized models for each component of our pipeline, as shown in Table~\ref{tab:model_selection}. For query rewriting and response generation, we utilized Qwen3-235B-a22b due to its superior performance in understanding complex legal contexts and generating coherent responses. For legal literature filtering, we adopted Qwen3-14B to balance computational efficiency with filtering accuracy. The embedding and reranking tasks employed Qwen3-Embedding-8B and Qwen3-Reranker-8B, respectively, which were specifically designed for retrieval tasks.

\begin{table}[h]
\centering
\caption{Model selection for different stages}
\label{tab:model_selection}
\resizebox{\columnwidth}{!}{%
\begin{tabular}{lll}
\hline
\textbf{Subtask} & \textbf{Module} & \textbf{LLM} \\ \hline
\multirow{5}{*}{Conversational Retrieval} & Query Rewriting & Qwen3-235B-a22b \\
& Legal Literature Filtering & Qwen3-14B \\
& Dense Retrieval & Qwen3-Embedding-8B \\
& Literature Filtering & Qwen3-14B \\
& Reranking & Qwen3-Reranker-8B \\
\hline
Response Generation & Response Generation & Qwen3-235B-a22b \\
\hline
\end{tabular}%
}
\end{table}

\subsubsection{Hyperparameter Configuration}
For the multi-route retrieval system, we configured the dense retriever to recall 50 legal articles from each legal literature, while the sparse retriever (BM25) was set to retrieve 1,000 articles from the entire corpus. The weight parameter $\alpha$ for the dense retriever was set to 1.0, indicating complete reliance on dense retrieval for the final scoring. During the filtering stage, we processed the top 500 results from the hybrid retrieval to balance computational efficiency with recall performance. The final reranking stage selected the top 5 most relevant legal articles for response generation.

\subsubsection{Rerank Model Fine-tuning}
We fine-tuned the rerank model using the Swift framework. For the selection of training data, we used the annotations of legal provisions from user queries in the training data as positive samples. However, when selecting negative samples for training, we did not randomly pick legal provisions irrelevant to user queries. Instead, we randomly selected 5 provisions from the top 10 most relevant ones obtained after legal provision retrieval and filtering to serve as negative samples. It is worth noting that these 5 negative samples do not include the annotated positive samples.

\subsection{Ablation Study}
We conducted extensive ablation studies to validate the effectiveness of each component in our pipeline. Table~\ref{tab:ablation} presents the results of our ablation experiments across different configurations.

\begin{table}[h]
\centering
\caption{Ablation study results.}
\label{tab:ablation}
\begin{tabular}{lc}
\hline
\textbf{Configuration} & \textbf{Overall Score} \\ \hline
Vanilla RAG & 30.74 \\
+ Query Rewriting & 31.66 \\
+ Rerank & 33.60 \\
+ Multi-route Retrieval & 35.57 \\
+ Legal Literature Filtering & 36.89 \\
\hline
\end{tabular}
\end{table}

The results demonstrate that each component contributes positively to the overall performance. Query rewriting provides significant improvement in overall performance, highlighting the importance of contextualizing multi-turn conversations. The literature filtering mechanism effectively reduces noise in the candidate set, leading to improvements in both retrieval and generation metrics. The multi-route retrieval strategy combining dense and sparse methods shows modest but consistent gains across all metrics.

\subsection{Performance Analysis}
\begin{figure}[htbp!]
\centerline{\includegraphics[scale=0.25]{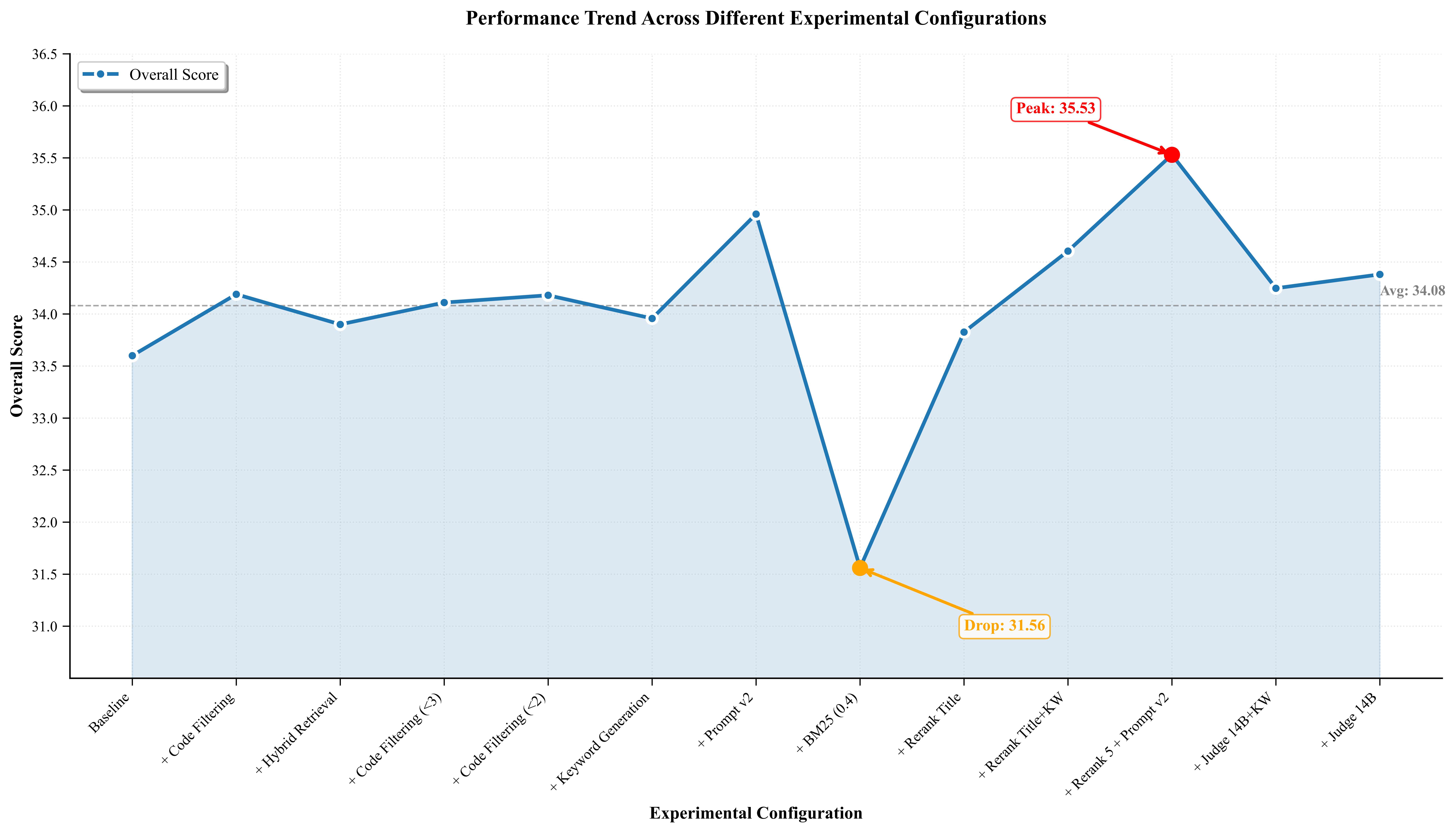}}
\caption{Performance trend across different experimental configurations. The x-axis represents different submission versions, and the y-axis shows the overall score.}
\label{fig:performance_trend}
\end{figure}

Our approach achieved competitive performance on both the validation and test sets. Extensive experiments on the CCIR CUP 2025 dataset demonstrate the effectiveness of our approach. Our system achieved a competitive score on the test set. Ablation studies confirm that each component contributes meaningfully to overall performance. Figure~\ref{fig:performance_trend} illustrates the performance evolution across different experimental configurations during our development phase.

The final system achieved a score of 39.38 on the test set, demonstrating good generalization capability. 

\subsubsection{Component Analysis}
We analyzed the performance contribution of each component:
\textbf{Query Rewriting:} The query rewriting module successfully addressed coreference resolution and topic drift issues in multi-turn conversations. Manual analysis of 100 randomly sampled conversations showed that 87\% of rewritten queries better captured user intent compared to using only the current turn's question.

\textbf{Literature Filtering:} The legal literature filtering stage effectively reduced the search space while maintaining high recall. On average, this stage filtered out 73\% of irrelevant legal codes while retaining 96\% of relevant ones, significantly improving downstream processing efficiency.

\textbf{Multi-route Retrieval:} The combination of dense and sparse retrieval methods provided complementary benefits. Dense retrieval excelled at capturing semantic similarity, while BM25 effectively handled exact keyword matches critical in legal contexts.

\textbf{Reranking:} The fine-tuned reranking model achieved the highest precision in the final ranking stage, with an average improvement of 0.08 in NDCG@5 compared to using retrieval scores alone.

\section{Case Analysis}
To better understand the effectiveness of our approach, we present detailed analyses of representative cases that demonstrate both the strengths and limitations of our system.

\subsection{Successful Cases}
\textbf{Case 1: Multi-turn Legal Consultation}

Consider a multi-turn conversation where a user initially asks: "Someone came to my house and hid..." and later follows up with: "But my brother knew he was committing murder. Can his behavior still be considered self-defense?" and finally:``My brother is only thirteen years old. Does he need to bear responsibility in this situation?''
Our system successfully handled this complex scenario through the following steps:
\begin{enumerate}[leftmargin=*]
\item \textbf{Query Rewriting:} The system correctly identified that the current question about age-related criminal responsibility needed to be contextualized with the previous discussion about self-defense and murder.
\item \textbf{Retrieval Results:} The system retrieved relevant articles, including Article 17 of the Criminal Law of the People's Republic of China (regarding age of criminal responsibility) and Article 20 (regarding self-defense).
\item \textbf{Generated Response:} The final response appropriately cited: "According to Article 17 of the Criminal Law of the People's Republic of China, persons under fourteen years of age are not subject to criminal responsibility. Therefore, your thirteen-year-old brother does not need to bear criminal responsibility in this situation."
\end{enumerate}
This case demonstrates our system's ability to maintain context across multiple turns while accurately identifying and retrieving relevant legal provisions.

\textbf{Case 2: Complex Legal Reasoning}

In another case involving vehicle installment payments and insurance requirements, our system successfully:
\begin{itemize}[leftmargin=*]
\item Identified the dual legal issues (contract law and consumer protection)
\item Retrieved Article 428 of the Civil Code regarding installment contracts
\item Generated a comprehensive response addressing both the payment obligations and insurance requirements
\item Maintained the professional legal consultation tone throughout the response
\end{itemize}

\subsection{Challenging Cases}
\textbf{Case 3: Ambiguous Legal Context}

Some cases revealed limitations in our approach. For instance, when users provided vague descriptions of legal situations without sufficient context, our query rewriting module occasionally over-interpreted the user's intent, leading to the retrieval of overly specific legal provisions that didn't fully address the broader legal context.

\textbf{Case 4: Emerging Legal Issues}

In cases involving recently enacted laws or legal interpretations not well-represented in the training data, our system showed reduced performance in both retrieval accuracy and response generation quality.

\subsection{Error Analysis}
Our analysis of system errors revealed three primary categories:

\begin{enumerate}[leftmargin=*]
\item \textbf{Context Misunderstanding (32\% of errors):} The query rewriting module occasionally misinterpreted user intent, particularly in cases with complex pronoun references or implicit legal relationships.

\item \textbf{Retrieval Gaps (28\% of errors):} Some relevant legal provisions were not retrieved due to vocabulary mismatches between user queries and legal text, despite semantic similarity.

\item \textbf{Generation Inconsistencies (40\% of errors):} The response generation module sometimes produced responses that, while legally accurate, didn't fully address the user's specific concerns or lacked appropriate legal citation format.
\end{enumerate}



These insights highlight both the practical utility of our approach and areas for future enhancement.

\section{Conclusion}
In this paper, we presented a comprehensive approach for legal knowledge retrieval and generation that addresses the unique challenges of conversational legal consultation. Our system integrates multiple specialized components, including query rewriting, literature filtering, multi-route retrieval, and reranking to create an end-to-end solution for legal question answering. 

Our work contributes to the growing body of research on applying AI to legal domains, with potential applications in legal education, public legal assistance, and professional legal research. By making legal information more accessible through conversational interfaces, such systems could help bridge the gap between complex legal knowledge and public understanding.

However, we acknowledge the importance of responsible deployment of AI systems in legal contexts. Our system should be viewed as a tool to assist legal professionals and educate users about legal concepts, rather than a replacement for professional legal advice. Future deployments should include appropriate disclaimers and safeguards to ensure users understand the limitations of AI-generated legal information.

\section{Limitations and Future Work}
Despite the promising results, several limitations remain that present opportunities for future research:
\begin{enumerate}[leftmargin=*]
\item \textbf{Domain Adaptation:} The current system's performance varies across different legal domains. Future work could explore domain-specific fine-tuning to improve consistency across all areas of law.

\item \textbf{Reasoning Complexity:} While our system handles straightforward legal queries well, it struggles with cases requiring complex multi-step legal reasoning or integration of multiple legal principles.

\item \textbf{Dynamic Knowledge Updates:} The static nature of our legal corpus limits the system's ability to handle queries about recent legal developments or emerging jurisprudence.

\item \textbf{Evaluation Metrics:} Current evaluation metrics focus primarily on retrieval accuracy and semantic similarity. Future work could develop more comprehensive evaluation frameworks that assess legal reasoning quality and practical utility.
\end{enumerate}

\printbibliography

\end{document}